\documentclass[notitlepage, twocolumn,floatfix,superscriptaddress]{revtex4-1}
\usepackage{graphicx, epsfig, bm, amsmath}
\usepackage{amssymb}
\usepackage{amsmath}
\usepackage{color}
\usepackage{wasysym}
\usepackage{savesym}
\usepackage{hyperref}

\def\barray{\begin{array}}
\def\earray{\end{array}}
\def\be{\begin{equation}}
\def\ee{\end{equation}}
\def\ben{\begin{equation} \nonumber}
\def\een{\end{equation}}
\def\ban{\begin{eqnarray*}}
\def\ean{\end{eqnarray*}}
\def\ba{\begin{eqnarray}}
\def\ea{\end{eqnarray}}

\def\({\left(}
\def\){\right)}
\def\half{{1\over2}}

\def\axion{\mathcal{X}}

\begin{document}

\title{Gauge-flation trajectories in Chromo-Natural Inflation}
\author{Peter Adshead}
\affiliation{Kavli Institute for Cosmological Physics,  Enrico Fermi Institute, University of Chicago, Chicago, Illinois 60637, U.S.A}
\author{Mark Wyman}
\affiliation{Kavli Institute for Cosmological Physics,  Enrico Fermi Institute, University of Chicago, Chicago, Illinois 60637, U.S.A}
\affiliation{Department of Astronomy \& Astrophysics, University of Chicago, Chicago, Illinois 60637, U.S.A.}

\begin{abstract}
We provide a detailed discussion of the multifield trajectories and inflationary dynamics of  the recently proposed model of Chromo-Natural inflation, which allows for slow roll inflation on a steep potential with the aid of classical non-Abelian gauge fields. We show that slow roll inflation can be achieved across a wide range of the parameter space. We demonstrate that Chromo-Natural Inflation includes trajectories that match those found in Gauge-flation and describe how the theories are related. 
\end{abstract}
\maketitle


Inflation \cite{Guth:1980zm,Linde:1981mu, Albrecht:1982wi} is remarkably successful at accounting for the primordial perturbations.  While it is a relatively simple matter to arrange for an inflationary phase in the early universe, the challenge is to make it end, and end in such a way that the resulting universe resembles our observed universe. Single field inflation achieves this by utilizing a scalar field slowly rolling on a flat potential. Inflation ends when the potential becomes too steep relative to its height and the slow roll conditions are violated. This occurs at a particular point on the potential, and thus one may think of the position of the field as a clock which tells the time before inflation ends. In order to get sufficient inflation and solve the flatness and horizon problems, the field must roll slowly. In canonical inflation, this translates into a requirement that the potential be nearly flat relative to its height. Maintaining slow roll requires that the curvature of the potential also be small. While it is a simple matter to write down suitable potentials which have the desired properties, these potentials require a high degree of fine tuning in order that they remain stable against radiative corrections. This is known as the `eta' problem. Natural Inflation solves this problem by using an axion \cite{Freese:1990rb, Adams:1992bn} whose potential is protected from such corrections by a shift symmetry. Unfortunately for Natural Inflation, matching cosmic microwave background observations requires the model to have a Planck-scale axion decay constant, $f$ \cite{Freese:2004un}. Such a setup seems to be difficult, if not impossible, to realize in string theory \cite{Banks:2003sx}. 

In our recently proposed theory of Chromo-Natural Inflation \cite{Adshead:2012kp}, we demonstrated that it may be possible to alleviate this problem by coupling the axion to non-Abelian gauge fields in a classical, rotationally invariant configuration.  The interactions between the axion and
the gauge fields generate a slowly rolling inflationary solution for a wide range of parameters, including
both large and small values for the axion decay constant. Importantly, this range includes values $f \ll M_{\rm pl}$. 

Our scenario differs in important ways from the many others that involve either axions, vector-like fields, or both. The seminal example of an axionic inflationary theory, 
Natural Inflation, requires $f \sim M_{\rm pl}$ as we noted before \cite{Freese:1990rb}. Within axionic theories, a variety of methods have been attempted to cure the need for Planckian decay constants, e.g. \cite{Kim:2004rp, Dimopoulos:2005ac, Easther:2005zr, Silverstein:2008sg, Germani:2010hd}. In a closer approach to our model, Ref.\ \cite{Anber:2009ua} uses the copious emission of Abelian gauge quanta to permit inflation with $f\ll M_{\rm pl}$. The authors of \cite{Gumrukcuoglu:2010yc,Kanno:2010nr,Watanabe:2009ct,Watanabe:2010fh, Yamamoto:2012sq} study models of inflation with a uniform gauge-kinetic coupling of the inflaton to multiple vector fields, while models of inflation where the curvature perturbations are partially produced by the vacuum fluctuations of a vector multiplet were considered by \cite{Dimopoulos:2008yv,Bartolo:2009pa,Bartolo:2009kg, Dimastrogiovanni:2010sm, Bartolo:2011ee}. A number of models have been proposed where gauge or vector-like fields have classical background values that play a central role in the inflationary mechanism (e.g. \cite{Ford:1989me,ArmendarizPicon:2004pm,  Koivisto:2008xf, Golovnev:2008cf,Golovnev:2009ks, Alexander:2011hz}), but wherein the inherent anisotropy of the vector fields is cured by invoking many such fields, rather than through the rotationally invariant background configuration we consider. We note that the models of \cite{Ford:1989me,ArmendarizPicon:2004pm,  Golovnev:2008cf,Golovnev:2009ks}, but not \cite{Alexander:2011hz}, were shown to be unstable in \cite{Himmetoglu:2008zp,Himmetoglu:2008hx,Golovnev:2009rm,Himmetoglu:2009qi}. More generally, other researchers have found that non-minimally coupled p-form fields can also generate inflationary backgrounds without fundamental scalar fields \cite{Koivisto:2009sd, Germani:2009iq, Germani:2009gg}; however, the stability of these models still appears to be an open question \cite{Golovnev:2009rm}. Finally, other work has demonstrated that self interacting 3-form fields can give rise to accelerating cosmologies  \cite{Koivisto:2009ew, Koivisto:2009fb}.

Classical, cosmological solutions for gauge fields have a history going back to the late 1970's, when there was a search for solutions to the Einstein-Yang-Mills field equations \cite{Cervero:1978db, Henneaux:1982vs, Hosotani:1984wj, Moniz:1990hf}. More recently, these configurations have been studied in the context of dark energy \cite{ Gal'tsov:2010dd, Galtsov:2011aa,Elizalde:2012yk} and it was noticed that such a configuration allows for inflation -- Gauge-flation -- without the presence of a scalar field \cite{Maleknejad:2011jw,Maleknejad:2011sq} (see also \cite{Galtsov:2011aa}). While such gauge field configurations have been studied in the context of inflating backgrounds before \cite{Moniz:1991kx}, in that earlier work the authors only studied situations in which the inflaton was charged under the gauge group but otherwise the gauge fields played no role in generating the inflationary epoch.

This paper is organized as follows. In Sec.\ \ref{sec:scni}  we provide a detailed description of the trajectories in Chromo-Natural Inflation.  In Sec.\ \ref{sec:params}, we describe the space of parameters which generate sufficient inflation. In Sec.\ \ref{sec:GFvsCNI} we elucidate the relationship between the model of Chromo-Natural Inflation \cite{Adshead:2012kp} and the model of Gauge-flation \cite{Maleknejad:2011jw,Maleknejad:2011sq}. We conclude in Sec.\ \ref{sec:concl}. Throughout this work, we use natural units where the reduced Planck mass $M_{\rm pl}= c = \hbar = 1$.

\section{The Space of Chromo-Natural Inflation}\label{sec:scni}

In our previous work \cite{Adshead:2012kp}, we proposed a model of inflation with an axion, $\axion$, and three non-Abelian SU(2) gauge-fields:
\begin{align}\label{eqn:action}\nonumber
\mathcal{L} = \sqrt{-g} & \bigg[-\frac{R}{2}-\frac{1}{4}F_{\mu\nu}^{a}F_{a}^{\mu\nu}  - \frac{1}{2}(\partial \axion )^2
\\ & - \mu^4\left(1+\cos\left(\frac{\axion }{f}\right)\right)-\frac{\lambda}{8f} \axion  F^{a}_{\mu\nu}\tilde F_{a}^{\mu\nu}\bigg],
\end{align}
where $F^{a}_{\mu\nu}$ is the usual non-Abelian gauge field strength tensor
\begin{align}
F^{a}_{\mu\nu} = \partial_{\mu}A^a_{\nu} - \partial_{\nu}A^a_{\mu} - \tilde g f^{a}_{bc}A^{b}_{\mu}A^c_{\nu},
\end{align}
$\tilde g$ is the gauge field coupling, $f^{a}_{bc}$ are the structure constants of SU(2), and $\tilde F_{a}^{\mu\nu} = \epsilon^{\mu\nu\alpha\beta}F_{a\alpha\beta}$. Following usual practice, Greek letters indicate spacetime indices while Roman letters represent gauge indices. 

We take the gauge fields to be in a classical configuration given by the ansatz, 
\begin{align}
A^{a}_{0} = 0,\quad A^{a}_i  =\psi(t) \,a(t)\delta^{a}_i.
\end{align}
That is, our gauge sector has a vacuum expectation value (VEV).
 This configuration of the gauge fields is notable because it leads to an isotropic and spatially homogeneous cosmological solution where isotropy is protected by the non-Abelian fields' gauge invariance -- rotations in real space are `undone' by rotations in gauge space, leaving the fields invariant. The SU(2) gauge symmetry is crucial here, as its global part can be associated with rotations in 3-space leading to the invariance.
 
\begin{figure*}[t] 
   \centerline{\psfig{file=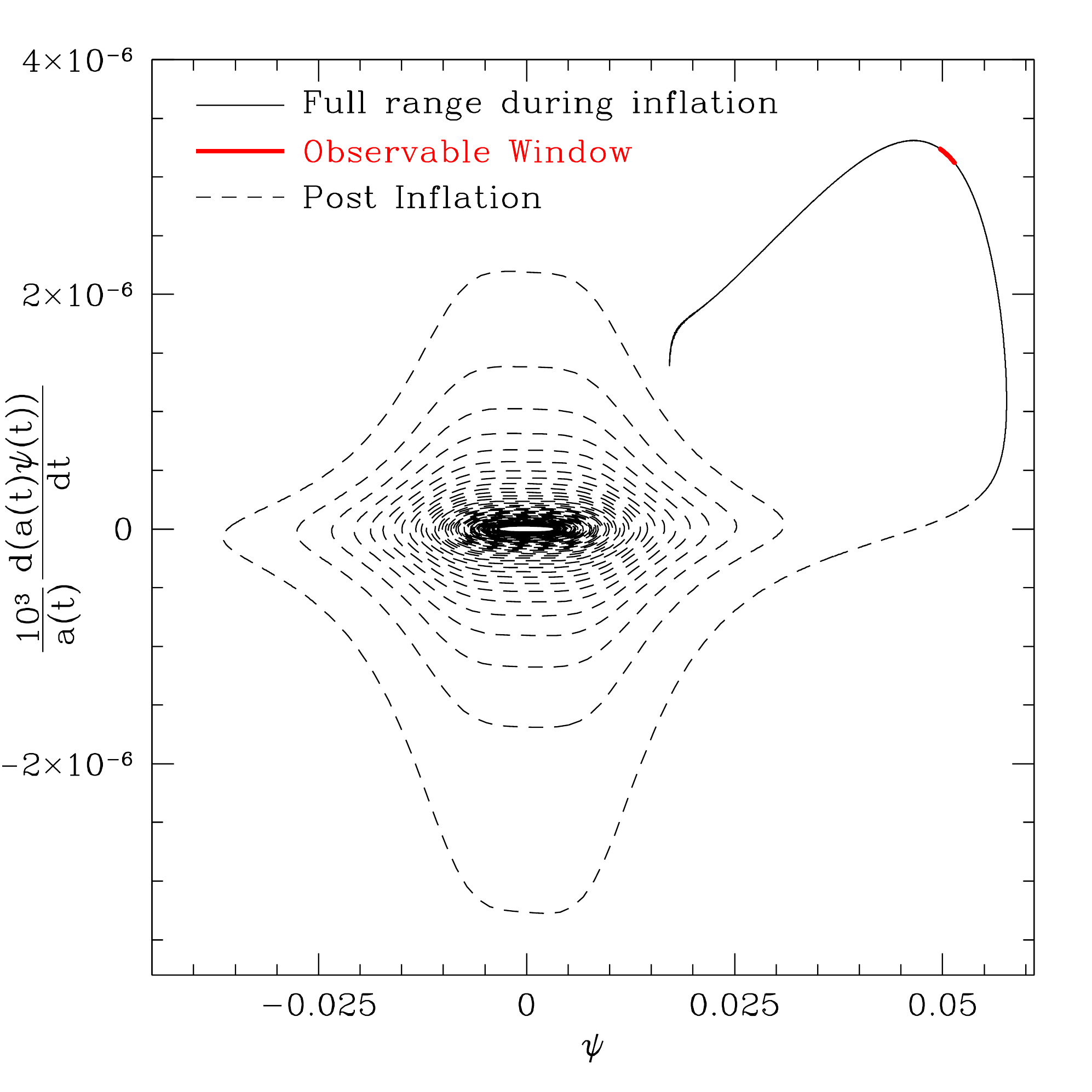, width=3.5in} \psfig{file=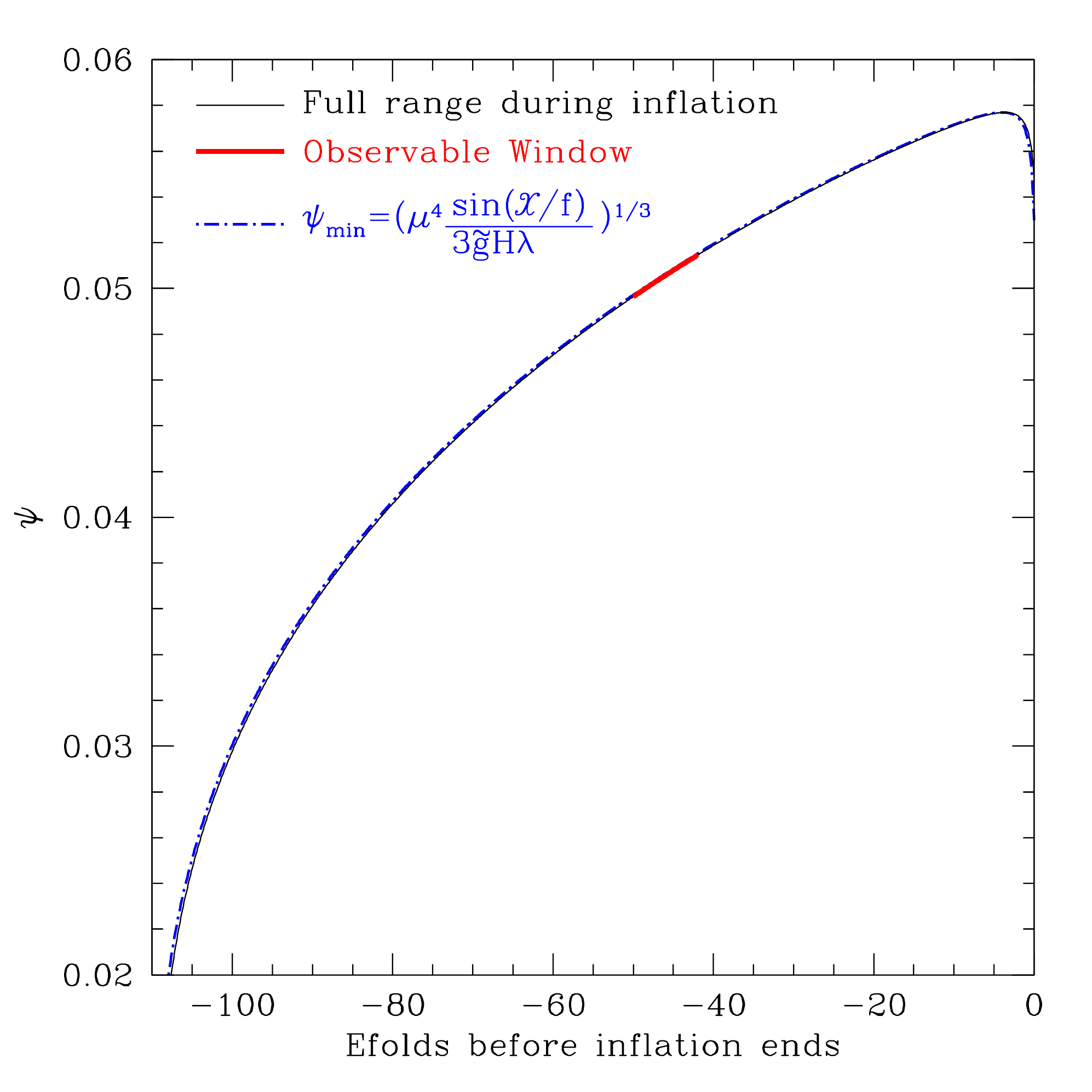, width=3.5in}}
   \caption{The behaviour of the gauge field during inflation for the choice of parameters $\{\mu,f,\tilde g, \lambda\} = \{3.16 \times 10^{-4}, 0.01, 2.0 \times10^{-6},  200\}$. In the left panel we show the phase portrait of the gauge field. The solid black curve corresponds to the period of exponential expansion, inflation. The red curve is the $\sim 8$ observable efoldings, 50 efoldings before the end of inflation, where the cosmic microwave background fluctuations are produced. Inflation ends where the dashed black line begins, and the gauge field decays. In the right hand panel we show the behavior of the gauge field as inflation proceeds, the observable window 50 efoldings before the end of inflation is shown in red. Plotted in blue is the value of the gauge field that minimizes its effective potential, Eqn.\ (\ref{eqn:psimin}).}
   \label{fig:Psi}
\end{figure*}

\begin{figure}[t*] 
   \centering
   \includegraphics[width=0.45 \textwidth]{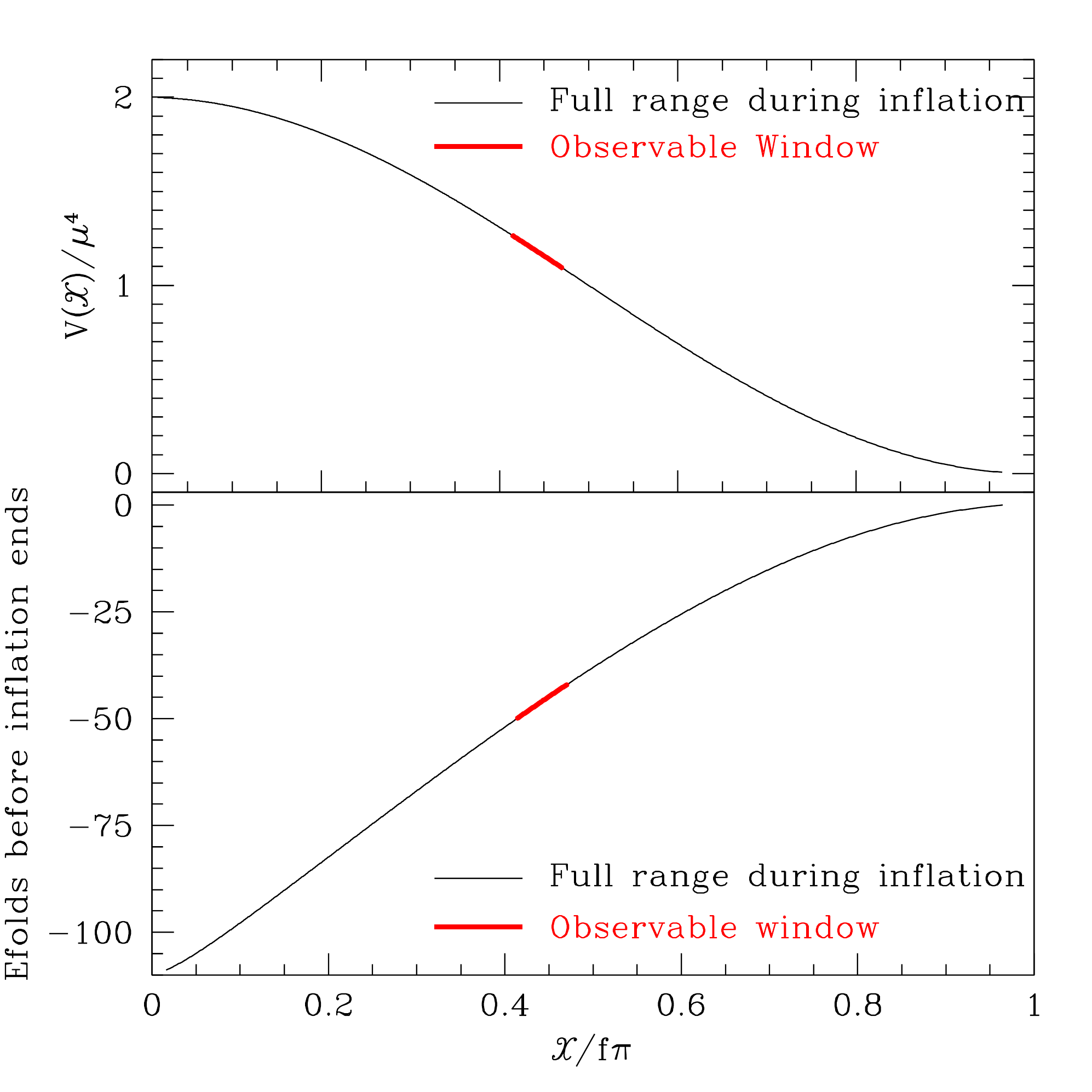}     
   \caption{The part of the bare axion potential traversed 50 efoldings before inflation ends (upper panel) is shown in red over the top of the parts of the potential that are probed by the axion during the entire period of inflation for the choice of parameters $\{\mu,f,\tilde g, \lambda\} = \{3.16 \times 10^{-4}, 0.01, 2.0 \times10^{-6},  200\}$. In the lower panel we show the full range of values the axion takes during inflation as a function of the efolding number. The  axion's position in between  50 - 42 efoldings before the end of inflation is shown in red.}
   \label{fig:axion}
\end{figure}

{ In the slow roll approximation, we can diagonalize the resulting equations of motion for the velocities, which gives
 \begin{align}
\label{Xdiag}
\Bigl(3H+\frac{g^2\lambda^2}{H f^2}\psi^4\Bigr)\dot\axion &= \frac{\mu^4}{f}\sin(\axion/f) - \frac{g\lambda}{f} H\psi^3 + \frac{2g^3\lambda}{ f H} \psi^5 \\
\label{psidiag}
\Bigl(3H+\frac{g^2\lambda^2}{H f^2}\psi^4\Bigr)\dot\psi &= -2H^2\psi - 2g^2\psi^3 - \frac{g^2\lambda^2}{f^2}\psi^5 \nonumber \\ 
& + \frac{g\lambda}{3H f^2}\psi^2\mu^4\sin({\cal X}/f) \ .
\end{align}
}
As was demonstrated in \cite{Adshead:2012kp}, the interacting system of a scalar field and non-Abelian gauge fields described by the action in Eqn.\ (\ref{eqn:action}) and the equations of motion, Eqns. (\ref{Xdiag}) and (\ref{psidiag}), is extremely effective at generating slow roll inflation in the presence of steep bare scalar potentials.  This is because the gauge field VEV is dynamically forced to a trajectory where it generates a very flat potential for the scalar {since the scalar feels an extra damping from the gauge field.} In our case, where the potential is assumed to be a cosine, this trajectory is very accurately given by (see Fig.\ \ref{fig:Psi})
\begin{align}\label{eqn:psimin}
\psi_{\rm min} = \left(\frac{\mu^4\sin\(\frac{\axion}{f}\)}{3\tilde g \lambda H}\right)^{1/3}.
\end{align}
On this trajectory, the change in potential energy as the axion rolls is almost entirely converted into gauge field energy rather than the axion's kinetic energy. This means that the axion rolls only slowly on its otherwise steep potential. {The gauge field's VEV, in turn, is being classically sourced by the axion's roll, which is why the VEV remains approximately constant despite
the presence of an exponentially expanding space \footnote{In our previous work, \cite{Adshead:2012kp}, we considered the question of whether the emission of gauge quanta, which happens in addition to the classical generation of the gauge field, will spoil our set-up. A rough calculation, found in a footnote there, suggests that the quantum emission is far smaller than
the classical generation. Additionally, we note that the gauge field gets a dynamical mass around the minimum given in the text, Eqn. \ref{eqn:psimin}, of ${\cal O}(H)$. This effective mass will
further suppress the emission of gauge quanta beyond the calculation that we mention in the previous work.}.}
This solution makes no assumptions about where the axion is on its potential. The only assumption needed is that we have chosen parameters such that 
\be 
3f^{2}H^2 \ll \tilde g^2\psi^4 \lambda^2,
\ee which amounts to choosing $\lambda \gg 1$. For our purposes, $\lambda \sim \mathcal{O}(100)$ is sufficient. This permits us to achieve observationally
viable inflation with $f \ll M_{\rm pl}$. A typical inflationary path for $\axion$ is illustrated in Fig. \ref{fig:axion}. Note that inflation proceeds over nearly the entire range of the axion's potential.

\begin{figure*}[t] 
   \centerline{\psfig{file=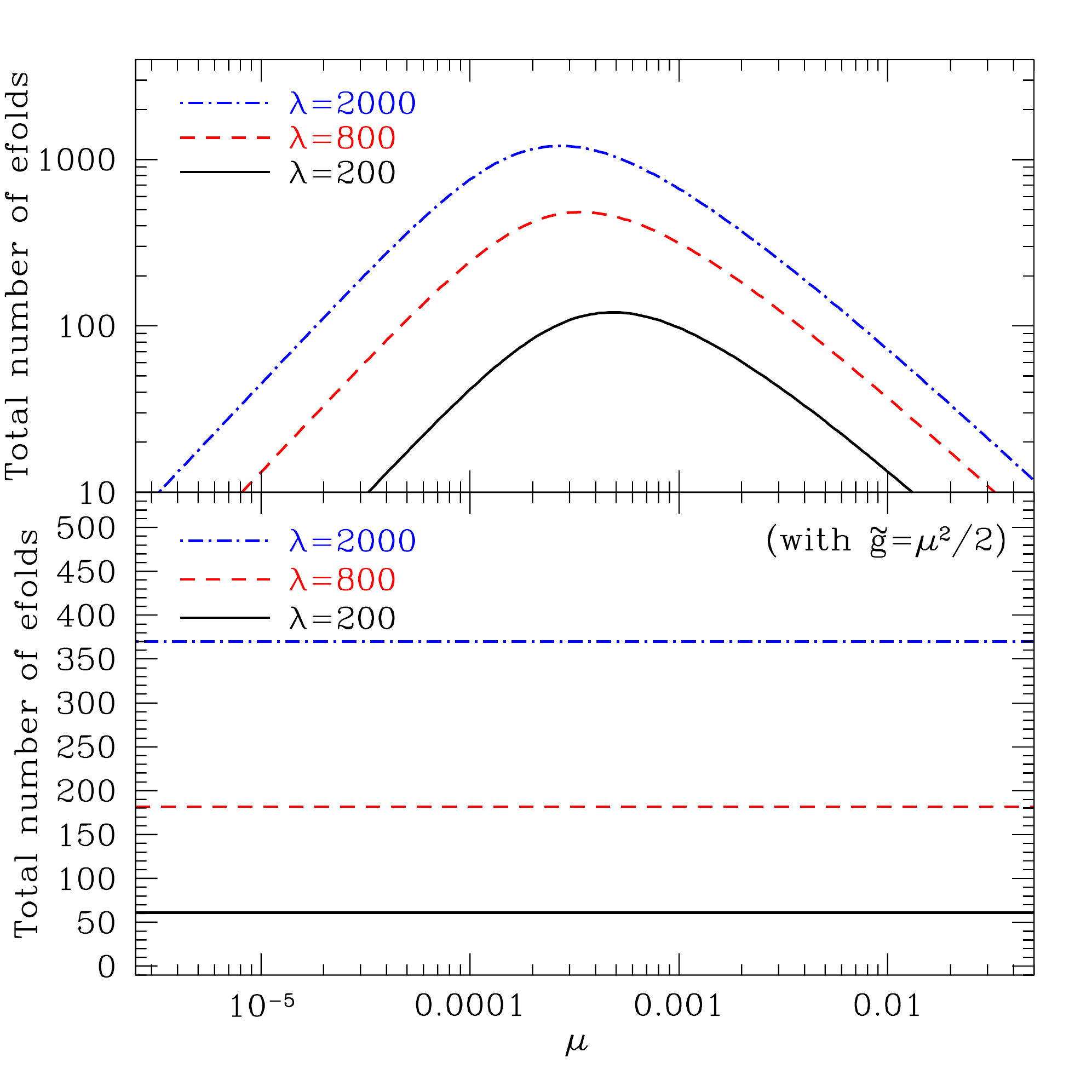, width=3.5in} \psfig{file=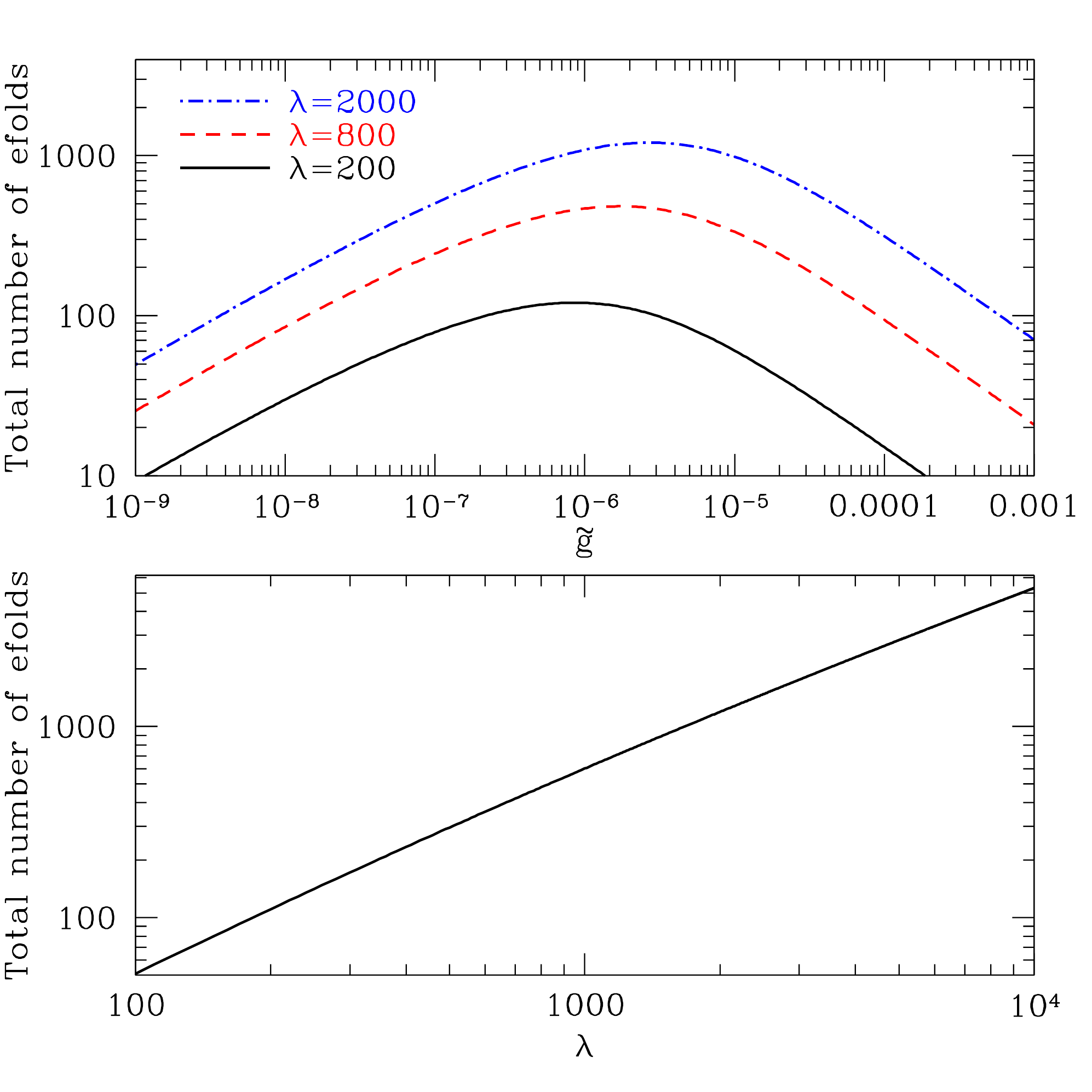, width=3.5in}}
   \caption{We show the number of efoldings of inflation produced as the various parameters of Chromo-Natural Inflation are varied. In the upper left panel, we show how the total amount of inflation varies as we vary the energy scale of the axion's potential, $\mu$. In the upper right panel we show how the total amount of inflation varies as we vary the gauge field coupling strength $\tilde g$. In the lower left panel, we show that the number of efoldings of inflation is kept constant if the parameters gauge field strength and the axion energy scale are covaried while keeping the ratio $2\tilde g/\mu^2$ constant. We also show the effect of these variations at various values of the coupling strength between the gauge and axion sectors, $\lambda$. In the lower right panel, we show the effect of varying $\lambda$ while keeping the remaining parameters fixed.    Unless otherwise noted, all other parameters are fixed at the values $\{\mu,f,\tilde g, \lambda\} = \{3.16 \times 10^{-4}, 0.01, 2.0 \times10^{-6},  200\}$. {This parameter set is one that we estimate will give the appropriate level of cosmological perturbations. In this figure, however, deviations from this set are not guaranteed to generate the appropriate level of perturbations, and generically will not.}}
   \label{fig:params}
\end{figure*}

It is instructive to ask what brings about the end of inflation. With this goal in mind, we can calculate the slow roll parameter $\epsilon_H =-\dot{H}/H^2$, where here and throughout, and overdot denotes derivatives with respect to cosmic time; inflation ceases when $\epsilon_H = 1$, and \cite{Adshead:2012kp}
\begin{align}\label{eqn:epsilon}
\epsilon_{H} \approx \frac{3\tilde g^2 \psi^4}{\mu^4\(1+\cos\(\frac{\axion}{f}\)\)}+\psi^2.
\end{align} 
This equation has a simple interpretation. The second term ($\psi^2$) is unimportant at the end of inflation and can be neglected. The numerator of the first term is related to the gauge field energy density, while the denominator is the axion's potential energy. Hence, inflation ends when the gauge field energy density becomes comparable to the axion's potential energy; in other words, when the Universe becomes radiation dominated rather than vacuum energy dominated. Since the gauge field energy is small and nearly constant throughout inflation, this happens when the axion reaches the bottom of its potential. 

In Chromo-Natural Inflation we are not required to make any assumptions about region of the potential where the axion is located -- we can simply evolve the system over the entire range,
$0 < \axion/f < \pi.$  Precisely how close the system can get to $\axion/f = \pi$ is determined by the first term in Eqn.\ (\ref{eqn:epsilon}). 

The system of equations corresponding to Chromo-Natural Inflation (see \cite{Adshead:2012kp})
is straightforward to solve numerically. For the plots in Figs.\ \ref{fig:axion} and \ref{fig:Psi}, we work with the parameter set  from \cite{Adshead:2012kp}
\be
\label{eqn:params}
\{\mu,f,\tilde g, \lambda\} = \{3.16 \times 10^{-4}, 0.01, 2.0 \times10^{-6},  200\},
\ee
and take the following initial conditions. The axion begins at $\axion(t_0) = 5\times10^{-4}$, with a velocity given by
\begin{align}
\left.\frac{d \axion}{dt}\right|_{t_{0}} = -\frac{\lambda}{f}\tilde{g}\dot{\psi}\psi^2.
\end{align}
For convenience, we initialize the gauge field at its attractor value,
\begin{align}
\psi(t_0) = \left(\frac{\mu^4\sin\(\frac{\axion(t_0)}{f}\)}{3\tilde g \lambda H}\right)^{1/3}.
\end{align}
However, let us emphasize that this is not necessary. If the gauge field starts away from this value, it relaxes to it within a few efolds. 
For similar reasons, we choose for the gauge field to have a small initial velocity, 
\begin{align}
\frac{\dot\psi}{H} = -1\times10^{-6}.
\end{align}

In Figure \ref{fig:axion} we show the region of field space over which the axion ranges during all of the inflationary period. The upper panel illustrates the potential, the lower panel the position of the axion as a function of the time before inflation ends. In both Figures \ref{fig:axion} and \ref{fig:Psi}, areas where the line is colored red indicate the region where the observable fluctuations are produced.

In the left panel of Figure \ref{fig:Psi}, we show the phase space of the gauge field. We show in solid lines the region that corresponds to the inflationary epoch, in red is the region where the fluctuations are produced, 50 efoldings before inflation ends. We also show, in dashed lines, the post inflationary evolution of the gauge field. Notice that the end of inflation and the post inflationary epoch exhibit similar behavior to that noted by the authors of \cite{Maleknejad:2011jw,Maleknejad:2011sq}. This period corresponds to the decay of the gauge field VEV and would likely provide a mechanism for reheating.

In the right hand panel of Figure \ref{fig:Psi}, we show the evolution of the gauge field as a function of the number of efoldings before inflation ends. As with the left hand panel, the observable range is shown in red. We also plot the curve corresponding to the condition in Eqn.\ (\ref{eqn:psimin}), noting the excellent agreement throughout the inflating epoch.

\begin{figure*}[t] 
   \centerline{\psfig{file=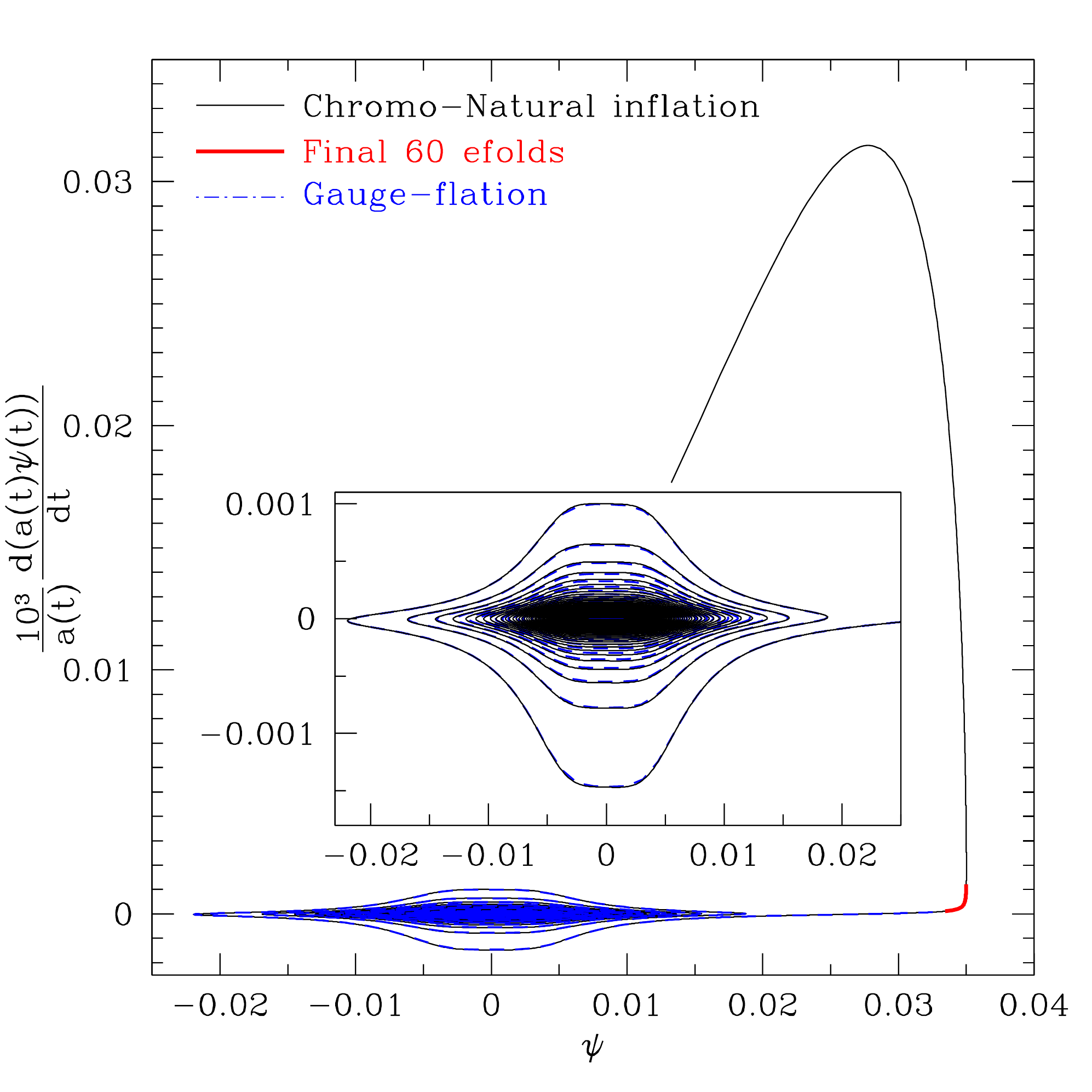, width=3.5in} \psfig{file=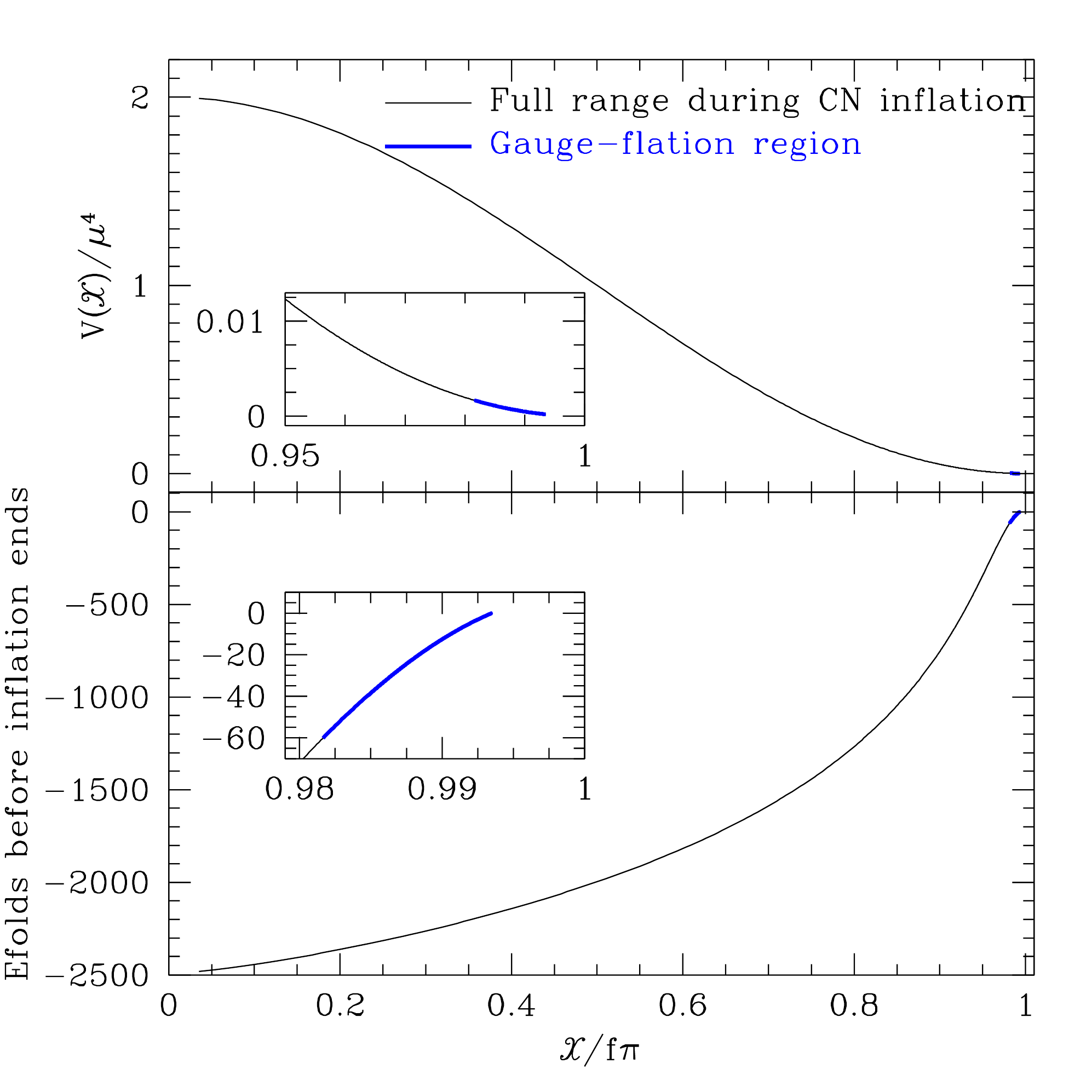, width=3.5in}}
   \caption{The behaviour of the gauge field during inflation for the choice of parameters that corresponds to the Gauge-flation model of \cite{Maleknejad:2011jw} $\{\mu,f,\tilde g, \lambda\} = \{4 \times 10^{-2}, 0.01, 2.5 \times10^{-3},  12158\}$. In the left panel we show a phase portrait of the gauge field. The solid black curve corresponds to the period of exponential expansion, inflation. The red curve shows the final $\sim 60$ efoldings of inflation and Chromo-Natural inflation ends at the end of the red curve here where the gauge field decays. The dashed blue line here (which begins 60 efoldings before the end of inflation, and thus includes the red curve) shows the result of evolving the equations that follow from the action after the axion has been integrated out, Eqn.\ (\ref{eqn:actionGF}), where $\{\tilde g, \kappa\} = \{2.5\times10^{-3}, 1.73\times 10^{14}\}$.  In the inset panel we show in more detailed the post-inflationary region.
In the right hand panel we show the full range of the axion as Chromo-Natural inflation proceeds. In blue we show the region that corresponds to the gaugeflation regime, which for these parameters is occurring for 60 efoldings. In the inset panels we show the region of the evolution of the axion where the two theories overlap.}
   \label{fig:CNIvsGF}
\end{figure*}

\section{The parameter space of Chromo-Natural Inflation}\label{sec:params}

We have so far focussed on the parameter set in Eqn. (\ref{eqn:params}) because we were able to show that it gives
an observationally viable amplitude and running of the spectrum of adiabatic density perturbations, at least 
at the level of approximation that was used in \cite{Adshead:2012kp}. Our estimate was based on the fact that the axion's position on its potential sets the `clock' for our model, as is the case in most single field inflationary models.  
Curvature fluctuations in inflation are due to spatial fluctuations in the time at which inflation ends. Our estimates were made assuming that the curvature fluctuations are generated by quantum fluctuations of the axion along its effectively flat direction. The spatial variation of the fluctuations translate into spatial fluctuations of the clock and hence produce a shift in the time inflation ends from place to place - a curvature fluctuation. We stress that this kind of calculation can only give rough estimates of the amplitude and tilt of the fluctuations. {From \cite{Adshead:2012kp}, this calculation gives
\be
{\cal R} \simeq \frac{1}{2 \pi} \frac{H}{\axion'} \sim \frac{1}{10} \frac{\lambda \mu^2}{f} \sim 10^{-4}
\ee
where $\cal R$ is the curvature perturbation, $\axion'$ is the axion's velocity in e-folding time, and in the final $\sim$ equivalences we approximate $\axion' \sim f/\lambda$, $H \sim \mu^2$.  Of course, a full analysis of all of the degrees of freedom of the theory is necessary to go beyond this estimate. Preliminary results suggest that the full story may be somewhat more complicated \cite{precite}.}

We do not expect this particular choice to exhaust the possible parameter space of viable models. Hence, in this section we explore the parameter space of models which generate sufficient inflation, and therefore solve the horizon and flatness problems. We have verified directly that it is very easy to achieve sufficient inflation ($N_{\rm efolds}>60$) in our model; all that is required is that our parameter $\lambda \gtrsim 100$, and of course one needs to start with the axion far enough up its potential. For our fiducial parameter set, $\axion_0 /(f\pi) < 0.4$ is sufficient. Any larger value of $\lambda$ will easily generate enough inflation, with mild assumptions about the other parameters (see Fig \ref{fig:params}). 

The behavior of the axion on its effective potential so well describes the trajectory of the axion that we can simply integrate our approximations to find the total number of efoldings that our model produces. The number of efoldings is found by integrating the expression \cite{Adshead:2012kp},
\be\label{eqn:efoldings}
N(\axion_0) = \int_{\frac{\axion_0}{f}}^\pi \hspace{-4 pt} \frac{\half \(3 \tilde  g^{2} \lambda ^{4} \tilde \mu^{4} \left( 1+\cos  x \right)^{2} \sin x\)^{1/3}}{ \( \lambda ^{2} \tilde \mu^{8} \left( 1+\cos x \right)^{4}\)^{1/3}+  \(3 \tilde g^{2} \sin x\)^{2/3} } dx,
\ee
over the entire range of the axions motion, $\axion/f \in (0, \pi)$. Unfortunately, this expression does not seem to admit any useful approximations which are valid over a large range of parameters, or at all points in the axion's range, and thus must be evaluated numerically. We note that this expression accurately matches the results from integrating
the full set of equations of motion across the full range of cases we consider.

In Fig.\ \ref{fig:params}, we illustrate the parameter space of Chromo-Natural Inflation. We show how the total amount of inflation varies as we move through the ranges of the parameters.

Notice that, if $\tilde g$ and $\mu$ are varied while keeping the ratio $\tilde g/\mu^2$ constant, the the total number of efoldings depends only on the value of $\lambda$. In the plot, we have chosen to hold $\tilde g/ \mu^2 = 1/2$ as an illustration of this effect.

It appears that the axion decay constant plays little to no role in this story.

\section{Relationship of Chromo-Natural Inflation and Gauge-flation}\label{sec:GFvsCNI}

References  \cite{Maleknejad:2011jw,Maleknejad:2011sq} propose the model of Gauge-flation, 
which is closely related to the one we have written down in  Eqn. (\ref{eqn:action}).
However, the authors of \cite{Maleknejad:2011jw,Maleknejad:2011sq} did not include the axion explicitly, but instead included a higher-order interaction
for the gauge fields:
\begin{align}\label{eqn:actionGF}
\mathcal{L_{\rm GF}} = \sqrt{-g} & \bigg[-\frac{R}{2}-\frac{1}{4}F_{\mu\nu}^{a}F_{a}^{\mu\nu} + \frac{\kappa}{384}   (F^{a}_{\mu\nu}\tilde F_{a}^{\mu\nu})^2\bigg].
\end{align}
Comparison between the models will immediately reveal  that this $(F^{a}_{\mu\nu}\tilde F_{a}^{\mu\nu})^2$ interaction term is precisely what one 
obtains upon integrating out the axion in our model. Integrating out a field can only be done when that field can be 
assumed to be in the minimum of its effective potential. Indeed, in \cite{SheikhJabbari:2012qf} it is explicitly shown that the Gauge-flation model corresponds to the Chromo-Natural Inflation model with the axion very near the bottom of its potential ($\axion \simeq \pi f$). In terms of the parameters of Chromo-Natural Inflation, 
\begin{align}
\kappa = 3 \frac{\lambda^2}{\mu^4},
\end{align}
and thus it follows that in the large axion regime ($\axion \simeq \pi f$), the theory of Chromo-Natural Inflation reduces to that of Gauge-flation. In order that such a model generates sufficient inflation, the axion-gauge field coupling $\lambda$ must be at least an order of magnitude larger than the minimum values needed in \cite{Adshead:2012kp}. This is hardly surprising, since one now needs to generate 60 efoldings of inflation while the axion is very near its minimum. For the choice of parameters from \cite{Adshead:2012kp} (Eqn.\ (\ref{eqn:params})), only 110 efoldings are generated in total and, as discussed above and illustrated in Fig. \ref{fig:axion}, during the required 60 efoldings the axion rolls more than half of the total distance in field space.
 
At the classical level, there is no impediment that we see to using this Gauge-flation parameter space to generate an inflating background solution.
However, it is also clear that this is a special case of the general model, Eqn. (\ref{eqn:action}). When the axion is included explicitly, one can describe cases far away from both the minimum, $\axion/f \simeq \pi$, and maximum, $\axion/f \simeq 0$,
of the axion's potential. In fact, given that one can always integrate out the axion near the bottom of its potential, Gauge-flation type trajectories
can always be matched to solutions of the full Chromo-Natural inflation model.

In the right panel of Fig.\ \ref{fig:CNIvsGF}, we demonstrate that the choice of parameters corresponding to the gaugeflation model of \cite{Maleknejad:2011jw} leads to more than 2500 efoldings of inflation in total in Chromo-Natural Inflation. The last 60 efoldings (shown in red) here lie well within the range where one can safely integrate out the axion to obtain the Gauge-flation model. The insets provide a more detailed view of the region where the theories overlap.

In the left hand panel of Fig.\ \ref{fig:CNIvsGF}, we show the full phase space of the Chromo-Natural Inflation trajectory. We also overplot the Gauge-flation result, which overlaps with the Chromo-Natural inflation result in the final 60 efoldings (shown in red here).  We have inset a zoomed in plot of the `reheating' phase to demonstrate the agreement in this region. As one would expect, the curves are virtually indistinguishable.

\section{Conclusions}\label{sec:concl}

In this work, we have described the space of trajectories in Chromo-Natural inflation in detail and demonstrated that sufficient inflation can be generated for a wide range of parameter values. We have shown that the Gauge-flation model of \cite{Maleknejad:2011jw} is subsumed by the more general model of Chromo-Natural inflation; Chromo-Natural inflation reduces to Gauge-flation when the axion is close to the minimum of its potential and can thus be integrated out. Integrating out the axion is an entirely valid way of simplifying the theory when the axion is near an extremum, and is a good way of describing the theory when the axion is near the minimum of its potential. In particular, one only recovers precisely the results of Gauge-flation once the axion nears the minima of its bare potential. 

This is not a surprising result. In the Gauge-flation model, the axion is essentially non-dynamical. It simply supplies the vacuum energy on which the universe inflates. While these trajectories are also described by Chromo-Natural Inflation, a much wider range of trajectories are also available where the dynamical behaviour of the axion over a large field range is important for the evolution of the system. Thus Chromo-Natural Inflation has a much larger model space. In particular, Chromo-Natural Inflation has two additional parameters compared to Gauge-flation, indicative of the greater freedom. A complete analysis of the perturbations in the various inflationary regimes \cite{precite} is required to discover if any are observationally viable,
and which observables might be able to discriminate among them.

\acknowledgements
We thank Richard Easther for comments and M.M. Sheikh-Jabbari for very helpful correspondence and for a sharing an early draft of his upcoming article. This work was supported in part by the Kavli Institute for Cosmological Physics at the University of Chicago through grants NSF PHY-0114422 and NSF PHY-0551142 and an endowment from the Kavli Foundation and its founder Fred Kavli.  MW was supported by U.S. Dept. of Energy contract DE-FG02-90ER-40560.

\bibliography{GFisCNI}

\end{document}